\documentclass{article}
\usepackage{graphicx,subfigure,amsmath} 
\usepackage{INTERSPEECH2020}

\title{Weakly Supervised Construction of ASR Systems with Massive Video Data}
\name{Mengli Cheng$^{\star\#}$, Chengyu Wang$^{\star\#}$, Xu Hu$^{\dagger\#}$, Jun Huang$^{\star\#}$, Xiaobo Wang$^{\star}$\thanks{$^{\#}$ M. Cheng and C. Wang contributed equally. J. Huang is the corresponding author. Work was conducted when X. Hu was with Alibaba Group.}}
\address{$^{\star}$ Alibaba Group $^{\dagger}$ ByteDance Inc.}

\begin{document}

\maketitle
\begin{abstract}
Building large-scale Automatic Speech Recognition (ASR) systems from scratch is significantly challenging, mostly due to the time-consuming and financially-expensive process of annotating a large amount of audio data with transcripts. Although several unsupervised pre-training models have been proposed, applying such models directly might be sub-optimal if more labeled, training data could be obtained without a large cost. In this paper, we present a weakly supervised framework for constructing ASR systems with massive video data. As videos often contain human-speech audio aligned with subtitles, we consider videos as an important knowledge source, and propose an effective approach to extract high-quality audio aligned with transcripts from videos based on text detection and Optical Character Recognition. The underlying ASR model can be fine-tuned to fit any domain-specific target training datasets after weakly supervised pre-training. Extensive experiments show that our framework can easily produce state-of-the-art results on six public datasets for Mandarin speech recognition.~\footnote{The pre-trained models, along with other resources will be released upon paper acceptance.}
\end{abstract}

\section{Introduction}

Automatic Speech Recognition (ASR) is one of the core tasks in speech processing, which aims to generate transcripts from speech utterances. Recently, end-to-end ASR models have been extensively studied, as these models do not require the explicit learning of acoustic and language models~\cite{DBLP:conf/icassp/ChanJLV16,DBLP:conf/icassp/SaonTAK19}.

Despite the success, a potential drawback is that these models require large amounts of transcribed data to produce satisfactory results~\cite{DBLP:conf/icassp/HsuCL20}. Unfortunately, transcribing audio by human annotators is both time-consuming and financially-expensive~\cite{DBLP:conf/tsp/ManolacheGCC20}. Recently, unsupervised pre-training has been applied to ASR~\cite{DBLP:conf/interspeech/SchneiderBCA19,DBLP:journals/corr/abs-2006-11477}, which uses unlabeled audio to pre-train ASR models. However, it is difficult for these models to outperform 
semi-supervised or supervised approaches. A few methods generate syntactic speeches~\cite{DBLP:conf/icassp/ChenYYJS20,DBLP:conf/icassp/RossenbachZSN20}, but the generated speeches may still be different from real ones.

\begin{figure}
	\centering
	\includegraphics[width=0.85\columnwidth]{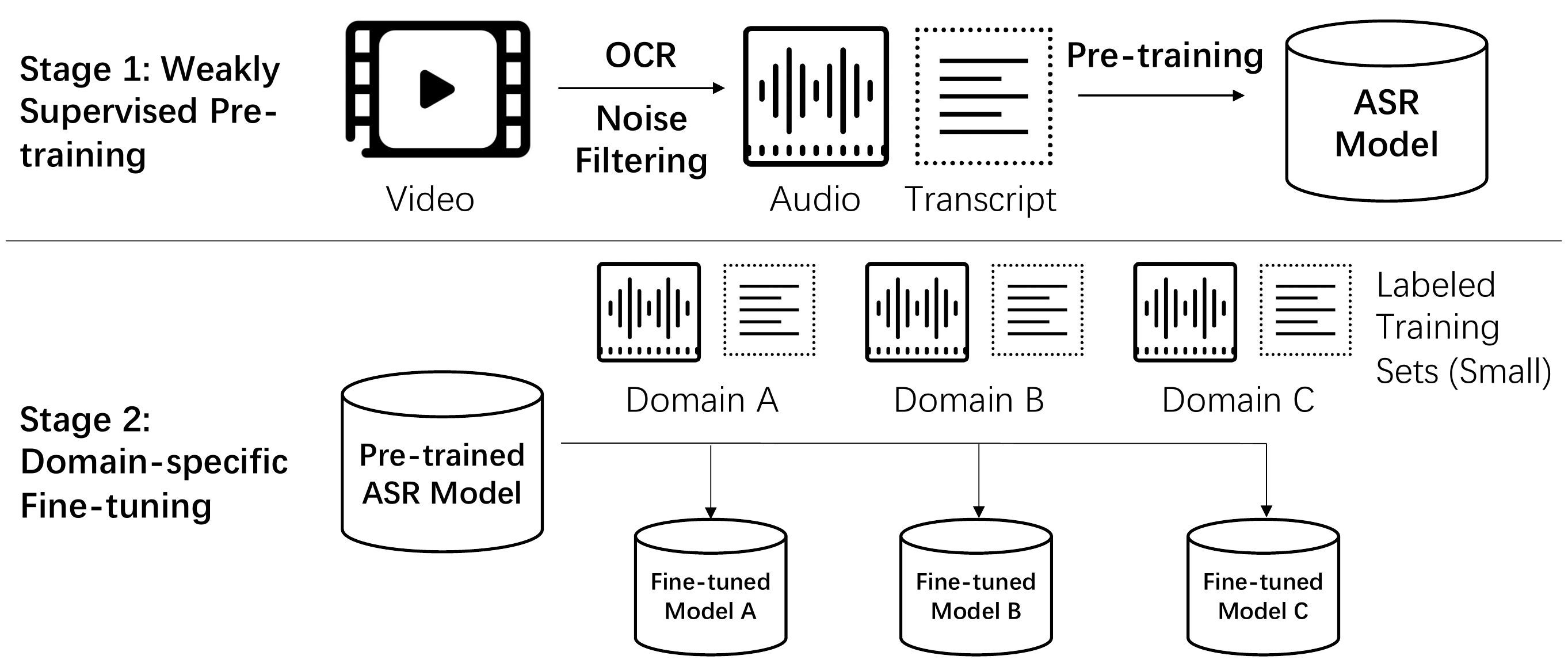}
 	\caption{High-level architecture of the proposed approach. }
	\label{fig:framework}
	\vspace{-.5em}
\end{figure}

A natural question arises:~\emph{is it possible to build accurate end-to-end ASR systems without much labeled data?} Here, we present a weakly supervised framework to construct ASR systems from massive video data, shown in Figure~\ref{fig:framework}. It consists of two major stages:~\emph{Weakly Supervised Pre-training} (WSP) and~\emph{Domain-specific Fine-tuning} (DF). During WSP, based on text detection~\cite{DBLP:conf/ijcai/YangCZCQL18} and Optical Character Recognition (OCR)~\cite{DBLP:conf/cvpr/LeeO16}, we extract human-speech audio aligned with subtitles from videos as knowledge sources to pre-train ASR models.
We pre-train our models over video of varied topics so that pre-trained models can capture transferable, general knowledge across domains. After that, the underlying ASR models can be fine-tuned to fit training data (usually smaller in size) in any domains. The framework is highly general as it can be applied to arbitrary languages and any end-to-end ASR models.  We evaluate our framework over popular ASR models and public datasets. Results show that it produces state-of-the-art results for Mandarin speech recognition.

\section{Related Work}


\noindent\textbf{End-to-end ASR.}
While hybrid ASR techniques are continuously developing (such as classical DNN-HMM-style models~\cite{DBLP:conf/interspeech/TanakaMMOA19}), due to the simple model pipelines, end-to-end ASR models have gained much attention. 
Recurrent-style networks are naturally suitable for end-to-end ASR as they model the sequences of audio and languages~\cite{DBLP:conf/interspeech/PeddintiPK15,DBLP:conf/icassp/ChanJLV16,DBLP:conf/interspeech/PoveyPGGMNWK16}; however, they may be slow during training and inference. This reduces the application scopes of such models in industry. CNN-based approaches~\cite{DBLP:journals/corr/CollobertPS16} are faster in speed, but they have limited capacity for modeling long sequences.
Transformer-based methods~\cite{DBLP:conf/icassp/ZhaoLWL19,DBLP:conf/aaai/LiLW0ZL20,DBLP:conf/icassp/MoritzHR20,DBLP:conf/interspeech/HanHTHZ19} have better performance because have strong abilities to capture long-term dependencies. They also converge faster and produce more accurate results when the CTC (Connectionist Temporal Classification) loss is added as an auxiliary loss~\cite{DBLP:conf/icassp/MiaoCGZ020}. Because the architecture design is not our major focus, we do not further elaborate.

\textbf{Pre-training ASR Models.}
Two streams of works have been proposed to reduce the requirements of manually labeled data for end-to-end ASR. 
One stream applies unsupervised/semi-supervised methods to tackle the problem. For example, Long et al.~\cite{DBLP:journals/access/LongLWZY19} propose semi-supervised training of DNN and RNN based acoustic models. 
Inspired by BERT~\cite{DBLP:conf/naacl/DevlinCLT19}, 
Baevski et al.~\cite{DBLP:journals/corr/abs-2006-11477} and Schneider et al.~\cite{DBLP:conf/interspeech/SchneiderBCA19} propose masked predictive coding for unsupervised pre-training of transformer encoders. Although these methods can improve the performance, there is still performance gap from supervised model training with labeled data~\cite{DBLP:conf/interspeech/SchneiderBCA19,DBLP:journals/corr/abs-2006-11477}.

The other stream extracts aligned text-speech segments using existing ASR models.
Lanchantin et al.~\cite{DBLP:conf/interspeech/LanchantinGKLQW16} align paragraphs of transcripts with audio to generate training data. 
The  works~\cite{DBLP:conf/asru/LiaoMS13,DBLP:conf/emnlp/LakomkinMWW18} introduce several heuristic rules to extract useful speech segments with transcripts from Youtube. However, these methods generally require a well-performed ASR model to start-up. Our work does not rely on accurate ASR models and can generate high-quality utterance-text pairs.


\section{The Proposed Framework}

In this section, we introduce technical details of our framework and the ASR model architectures that we use.

\subsection{Weakly Supervised Pre-training}

\begin{figure}
	\centering
	\includegraphics[width=0.9\columnwidth]{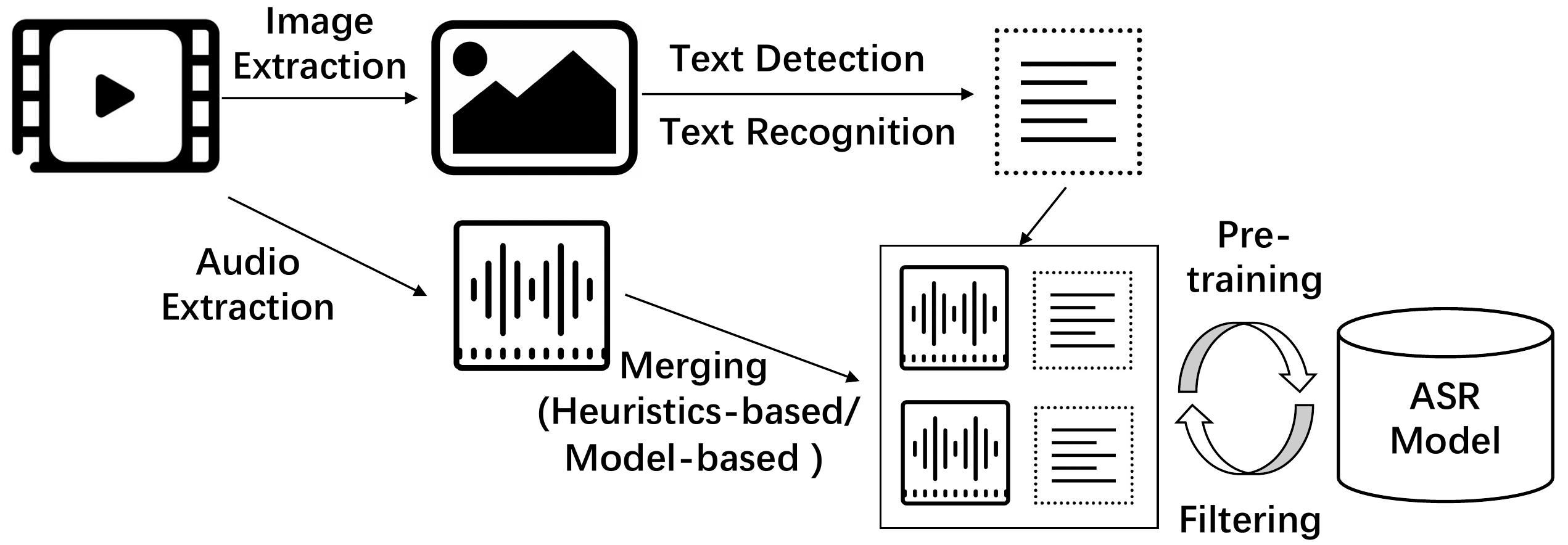}
 	\caption{Pipeline of Weakly Supervised Pre-training (WSP). }
	\label{fig:pipeline}
	\vspace{-.5em}
\end{figure}

The pipeline of WSP is illustrated in Figure~\ref{fig:pipeline}.

\noindent\textbf{Video Acquisition.}
Many videos have embedded subtitles that are almost synchronous with the audio. We regard such videos as pre-training knowledge sources. The videos that we use in this work are dramas of various genres provided from~\emph{Youku}~\footnote{\emph{Youku} (http://www.youku.com) is a popular video hosting service, a subsidiary of Alibaba Group. It holds the copyrights of these videos, and permits authors to obtain and process the data as described.}.

\noindent\textbf{Text and Audio Spotting.}
Although videos with subtitles are available to us, subtitles are generally embedded in frame images in different styles and formats, especially in videos made in early years. This prevents us from extracting subtitles from raw data sources directly. Hence, we first extract frame images from each video with an interval of 1/3 second. 
Next, we employ IncepText~\cite{DBLP:conf/ijcai/YangCZCQL18} to detect text positions from images and the OCR model~\cite{DBLP:conf/cvpr/LeeO16} to recognize text contents. 

Given a sequence of frame images within a time window size (denoted as $s_i, s_{i+1}, \cdots, s_{j-1}, s_j$), we wish to determine whether two consecutive frames $s_k$ and $s_{k+1}$ ($i\leq k$, $k+1\leq j$) can be ``merged'' so that a subset of such frames may correspond to the same subtitle. Hence, the audio within the time frames is treated as the speech for the subtitle. We present two merging methods:~\emph{Heuristics-based} and~\emph{Model-based}. For two consecutive frames $s_k$ and $s_{k+1}$, denote the detected texts as $t_k$ and $t_{k+1}$, respectively. Define the~\emph{Relative Edit Distance} (RED) between $s_k$ and $s_{k+1}$ as:
\begin{equation*}
RED(s_k, s_{k+1}) = \frac{EditDis(t_k, t_{k+1})}{\max(Len(t_k), Len(t_{k+1}))}
\end{equation*}
where $EditDis(t_k, t_{k+1})$ is the~\emph{edit distance} between $t_k$ and $t_{k+1}$, and $Len(t_k)$ is the length of $t_k$. \emph{Heuristics-based Merging} combines two frames $s_k$ and $s_{k+1}$ if $RED(s_k, s_{k+1})$ is smaller than a tuned threshold.

However, \emph{Heuristics-based Merging} ignores the corresponding relations between audio and texts. If any existing third-party ASR model is available, no matter whether it is accurate or not, we can use it to refine the merging process~\footnote{We use the model from~\emph{https://ai.aliyun.com/nls/asr}. The CER is slightly larger than 20\% based on their evaluation.}. Let $a_k$ be the audio segment w.r.t. the frame $s_k$. \emph{Model-based Merging} employs an existing model $f$ to predict the transcript of $a_k$, denoted as $f(a_k)$. If $s_k$ and $s_{k+1}$ should not be merged, the error rate of model $f$ is computed as:
\begin{equation*}
Err_1(f, s_k, s_{k+1})=CER(t_k, f(a_k))+CER(t_{k+1}, f(a_{k+1}))
\end{equation*}
where $CER(t_k, f(a_k))$ is the Character Error Rate (CER) of model $f$'s predictions. If $s_k$ and $s_{k+1}$ should be merged, similarly, we have the combined error rate:
\begin{equation*}
\begin{split}
Err_2(f, s_k, s_{k+1})=\min\{CER(t_k, f(a_{k:k+1})),\\
CER(t_{k+1}, f(a_{k:k+1}))\}
\end{split}
\end{equation*}
where $a_{k:k+1}$ concatenates $a_k$ and $a_{k+1}$. $s_k$ and $s_{k+1}$ should be merged if $Err_1(f, s_k, s_{k+1})>Err_2(f, s_k, s_{k+1})$.

\noindent\textbf{Iterative Pre-training.} 
After extraction and merging, we obtain a large ``pseudo-labeled'' dataset $D=\{(a_k, t_k)\}$, consisting of audio-transcript segment pairs.
Because supervised ASR model learning produces better results than unsupervised ones~\cite{DBLP:journals/corr/abs-2006-11477}, we pre-train the ASR model using the way as normal training over the dataset $D$. However, the extraction process of $D$ unavoidably injects noise into the dataset due to the lack of human annotation. During pre-training, we apply a self-training strategy to filter out noisy data. In each epoch, we filter out audio-transcript segments $\{(a_k,t_k)\}$ from $D$ that are most likely to have noisy transcripts and use the remaining dataset for the next training epoch. Due to space limitation, we omit the details and refer interested readers to~\cite{DBLP:conf/iccv/HuangQJZ19}.

\subsection{Domain-specific Fine-tuning}

Based on the WSP objective, our framework could generate ready-to-use ASR models directly. However, the domains of pre-training data may be significantly different from downstream ASR tasks. Hence, given a (small) training set $D_m=\{(a_k, t_k)\}$ of domain $m$, we fine-tune the pre-trained model over $D_m$ to learn domain-adaptive parameters. 

\subsection{Choices of Model Architectures}

\begin{figure}
	\centering
	\includegraphics[width=\columnwidth]{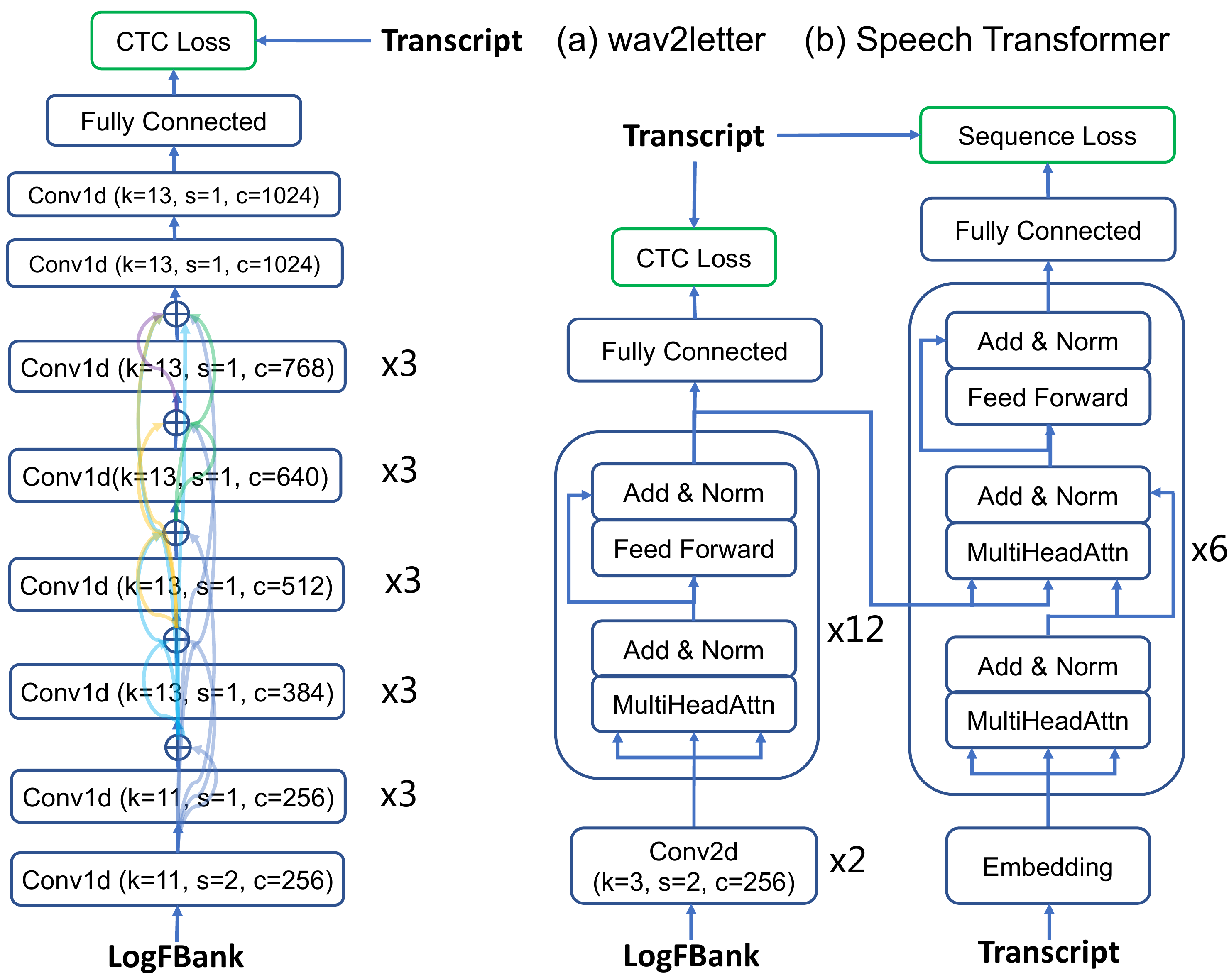}
 	\caption{The architectures of two models that we choose.}
 	\vspace{-.5em}
	\label{fig:models}
\end{figure}

Following industry practices, we consider two popular ASR models: wav2letter~\cite{DBLP:journals/corr/CollobertPS16} and Speech Transformer~\cite{DBLP:conf/icassp/ZhaoLWL19}, as shown in Figure~\ref{fig:models}. Wav2letter uses one dimensional convolution networks with large kernels as encoders, and the CTC loss for training. 
Its efficient inference speed makes it appealing to industrial applications.
Speech Transformer~\cite{DBLP:conf/icassp/ZhaoLWL19} adopts self-attention for acoustic modelling and decoding. Following~\cite{DBLP:conf/icassp/MiaoCGZ020}, the CTC loss is added to this model as an auxiliary loss to achieve faster convergence and better performance. In multi-head attention layers, we set the hidden size as 512, with 8 heads. We ensemble the last 10 checkpoints as our final model collection. For inference, we apply beam search of size 16 to all models in parallel to generate texts that are most probably correct.

\section{Experiments}

In this section, we conduct extensive experiments to evaluate the proposed framework in various aspects.

\begin{table}
\begin{footnotesize}
\begin{tabular}{llllll} 
\hline
\bf Dataset   & \bf Duration &  \bf SPK     &  \bf TXT    & \bf UTT         & \bf Style  \\ 
\hline
ST$\_$CMDS  & 500h   &   855    &  74,770  & 82,080       & R      \\
AISHELL-1 & 178h     &   400    &  113,738 & 120,099      & R      \\
AISHELL-2 & 1,000h    &   1,991   &  603,738 & 1,009,223     & R      \\
AIDATANG  & 200h     &   600    &  133,684 & 164,905      & R      \\
MagicData & 760h     &   1,080   &  275,778 & 573,480      & R      \\
HKUST     & 200h     &   2,100   &  173,028 & 173,028      & S     \\ 
\hline
\end{tabular}
\caption{Dataset statistics. (SPK: \#Speakers. TXT: \#Transcripts. UTT: \#Utterances. R/S: Reading/spontaneous style.)}
\label{tab:data_stat}
\end{footnotesize}
\end{table}

\subsection{Datasets and Experimental Settings}

We obtain 940 drama series under 16 categories from~\emph{Youku}, containing 43,694 video clips. The total duration is around 8,000 hours. During WSP, the learning rates of wav2letter and Speech Transformer are set as 0.05 and 1.0, respectively.
For both models, we normalize the utterances to 16kHz and generate the logarithm of FBank features of 80 dimensions, with a window size of 20ms and the stride of 10ms. We use our in-house IncepText and OCR models for text spotting. After WSP, we fine-tune and evaluate our models over six public datasets: ST$\_$CMDS\footnote{http://www.openslr.org/38/}, AISHELL-1\footnote{http://www.aishelltech.com/kysjcp/}, AISHELL-2\footnote{http://www.aishelltech.com/aishell$\underline{ }$2/}, AIDATANG\footnote{http://www.openslr.org/62/}, MagicData\footnote{http://www.openslr.org/68/} and HKUST\footnote{https://catalog.ldc.upenn.edu/LDC2005S15/}. The statistics are in Table~\ref{tab:data_stat}. The datasets are varied in domains and styles and have relatively short duration compared to our WSP dataset. During fine-tuning, we set the learning rates to be 0.01 and 0.5 for the two models. We keep the training, development and testing splits of all the datasets as default. All the algorithms are implemented in Tensorflow and run on GPU servers.

\begin{table}
\begin{footnotesize}
\begin{tabular}{l  c  c   c c c c }
\hline
\bf Model                                       & ST\_ & AS-1 & AS-2 & ADT & Magic &  HKUST  \\ 

& CMDS & & & & Data\\
\hline
TDNN~\cite{DBLP:conf/interspeech/PeddintiPK15}                   &   -      &    8.7    &    -      &   7.2    &    -      &  32.7   \\ 
Chain-Model~\cite{DBLP:conf/interspeech/PoveyPGGMNWK16}             &   -      &    7.5    &    -      &   5.6    &    -      &  28.1   \\
MS-Attn~\cite{DBLP:conf/interspeech/HanHTHZ19}                    &   -      &    -      &   8.5     &    -     &    -      &   -     \\ 
SpeechBERT~\cite{DBLP:journals/corr/abs-1910-11559}          &   -      &    7.4    &    -      &    -     &    -      &  21.0   \\
SAN-M~\cite{DBLP:journals/corr/abs-2006-01713} &   -      &    6.4    &    -      &    -     &    -      &  -   \\
\hline
wav2letter~\cite{DBLP:journals/corr/CollobertPS16}                                &  4.5     &    11.7   &   12.5    &  12.9    &   7.4     &  35.7   \\
wav2letter+WSP                    &  2.4     &    7.1    &   10.0    &  9.2     &   6.7     &  29.3   \\\hline
ST~\cite{DBLP:conf/icassp/ZhaoLWL19}                            &  4.4     &    6.7    &   7.4     &  7.8     &   3.6     &  23.5   \\
ST+WSP                     & \textbf{2.1} & \textbf{5.9}  & \textbf{5.9}  &  \textbf{4.9}  &  \textbf{3.3}  &  \textbf{20.0}   \\
\hline
\end{tabular}
\end{footnotesize}
\caption{Performance of ASR models on public test datasets in terms of CER (\%). ST refers to ``Speech Transformer'' with our modified architecture. AS-1 and AS-2 refer to ``AISHELL-1'' and ``AISHELL-2''.}
\vspace{-.75em}
\label{tab:all}
\end{table}

\begin{figure*}
	\centering
	\includegraphics[width=2.05\columnwidth]{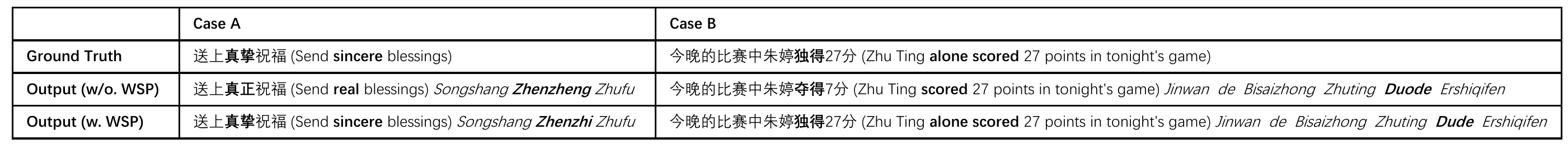}
    \vspace{-1em}
 	\caption{Cases of model prediction w. and w/o. WSP. Italic texts refer to pronunciation (spelled in Mandarin phonetic symbols).}
 	\vspace{-.5em}
	\label{fig:cases}
\end{figure*}

\subsection{General Performance Comparison}

We report the performance of our models in all test sets.  For baselines, we consider both classical and recent ASR models, including TDNN~\cite{DBLP:conf/interspeech/PeddintiPK15}, Chain-Model~\cite{DBLP:conf/interspeech/PoveyPGGMNWK16}, MS-Attn~\cite{DBLP:conf/interspeech/HanHTHZ19}, SpeechBERT~\cite{DBLP:journals/corr/abs-1910-11559} and SAN-M~\cite{DBLP:journals/corr/abs-2006-01713}. For wav2letter and Speech Transformer, we test the performance under both settings: i) w. WSP and ii) w/o. WSP. Results are summarized in Table~\ref{tab:all}. We have the following findings: i) Speech Transformer outperforms wav2letter across all the datasets\footnote{Despite its relatively high error rate, the wav2letter model still has 
wide applications in industry due to its simple architecture and fast inference speed. The applications are beyond the scope of this paper.}. ii) The WSP technique effectively boosts the performance of both models on all the datasets. This phenomenon is more significant on small datasets (i.e., AIDATANG and HKUST). iii) Speech Transformer with the WSP technique achieves state-of-the-art performance on all the six public datasets.

\subsection{Detailed Model Analysis}

\textbf{Analysis of WSP.}
To create the pre-training dataset, we test both merging techniques via a manual check on 0.2\% of the generated pairs. We observe that model-based merging produces better results. The CER is around 6\%, close to manually labeled datasets. 
This shows, even without human annotation, we can generate pre-training datasets with tolerable error rates. After text and audio spotting, we obtain a total of 1,825,927 utterances from all videos, ranging from 15-20s. 

Next, we evaluate the iterative pre-training technique. We filter out part of the data (quantified by the drop ratio $\gamma$) and take the rest as the pre-training data for the next iteration. We search for the best value of $\gamma$ from ${0, 0.5\%, 1.0\%, 2.0\%}$ and also compare our method with a classical data filtering approach~\cite{DBLP:conf/asru/LiaoMS13}. We use the third-party Mandarin ASR model from~\emph{https://ai.aliyun.com/nls/asr} for~\cite{DBLP:conf/asru/LiaoMS13}, instead of their original English ASR model.
In Table~\ref{tabl:data_filter}, we display the CER values produced by pre-trained wav2letter without fine-tuning, evaluated on the AISHELL-1 development set. It shows that WSP with $\gamma=1.0\%$ has the best performance.

\begin{table}[t!]
\begin{center}
\begin{footnotesize}
\begin{tabular}{llll}
\hline
\bf Method/Iteration & 4 & 8 & 12 \\
\hline
Liao et al.~\cite{DBLP:conf/asru/LiaoMS13} & 17.3 & 16.8 & 16.5\\
\hline
WSP ($\gamma=0$) & 16.1 & 15.0 & 14.2\\
WSP ($\gamma=0.5\%$) & 15.4 & 14.4 & 13.6\\
WSP ($\gamma=1.0\%$) & 15.3 & 14.2 & \bf 13.3\\
WSP ($\gamma=2.0\%$) & 15.6 & 14.9 & 14.7\\
\hline
\end{tabular}
\end{footnotesize}
\caption{Performance of pre-trained wav2letter with different data filtering techniques in terms of CER (\%).}
\label{tabl:data_filter}
\end{center}
\vspace{-1em}
\end{table}

\begin{figure}[htbp]
\centering
\subfigure[Model: wav2letter]{
\centering
\includegraphics[width=0.475\columnwidth]{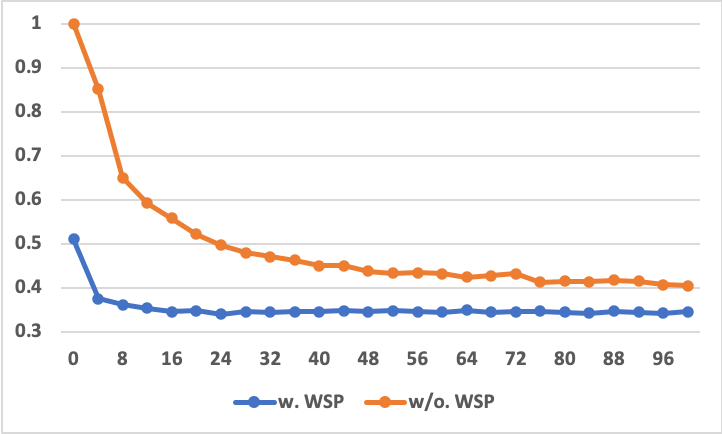}
}%
\subfigure[Model: Speech Transformer]{
\centering
\includegraphics[width=0.475\columnwidth]{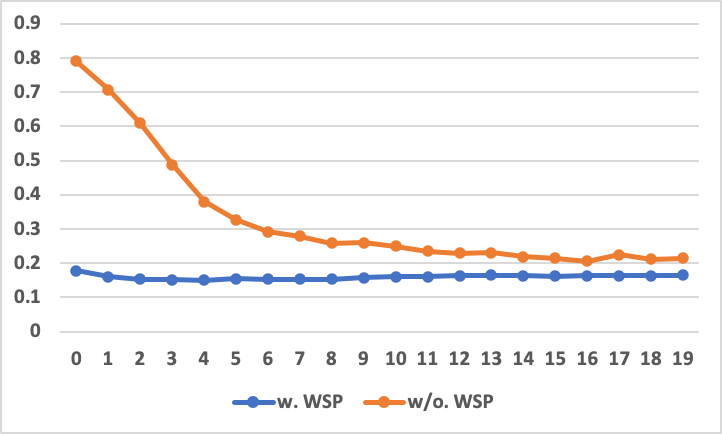}
}%
\centering
\caption{Convergence curves on HKUST. (X-axis: number of epochs; Y-axis: CER on the development set.)}
	\label{fig:train_curve}
\end{figure}

\noindent\textbf{Convergence analysis.}
We investigate how WSP affects the DF performance.
The convergence curves on HKUST are shown in Figure~\ref{fig:train_curve}. 
As seen, wav2letter and Speech Transformer converge within 10 and 3 training epochs, respectively. Compared to the same models without WSP, the speed of convergence is much faster for both models, which clearly indicates WSP is able to find better parameter initialization for domain-specific ASR tasks, no matter whether there exist domain differences between the two datasets.


\noindent\textbf{Error analysis and case studies.}
We analyze the percentages of different error types occurred in the test sets of AISHELL-1 and HKUST, shown in Table~\ref{tab:aishell_err}. The underlying ASR models are Speech Transformer w. and w/o. WSP.
The majority of the errors are substitution errors caused by homophones. The WSP technique helps to reduce such errors, as pronunciations and language contexts in the pre-training dataset are more diverse, leading to the better generalization ability of ASR models. Two typical cases can be also found in Figure~\ref{fig:cases}, with Chinese pronunciation and English translation provided. It shows WSP's ability to distinguish words with similar pronunciation.

\begin{table}[t!]
\begin{center}
\begin{footnotesize}
\begin{tabular}{lllll}
\hline
 \bf Dataset    & \bf WSP? & \bf Insertion & \bf Deletion & \bf Substitution \\
 \hline
 AISHELL-1    &    No    &    0.1    &  0.2     &   6.4           \\
 AISHELL-1    &    Yes   &    0.1    &  0.2     &   5.7        \\ \hline
 HKUST      &    No    &    2.6    &   3.6    &   17.3       \\
 HKUST      &    Yes   &    2.7    &   2.6    &   14.7       \\
 \hline
\end{tabular}
\caption{Error analysis in terms of CER (\%).}
\label{tab:aishell_err}
\end{footnotesize}
\end{center}
\vspace{-1.5em}
\end{table}

\section{Conclusion and Future Work}

In this paper, we construct accurate ASR systems based on weak supervision of massive video data. 
With WSP and the Speech Transformer model with our modifications, we achieve the state-of-the-art results on several public datasets. Future work includes i) applying our approach to other languages and ASR models; ii) combining unsupervised and weakly supervised pre-training in our framework; and iii) leveraging transfer learning to improve model fine-tuning.


\end{document}